\documentclass[10pt,journal,compsoc]{IEEEtran}

\ifCLASSOPTIONcompsoc
  \usepackage[nocompress]{cite}
\else
  \usepackage{cite}
\fi

\ifCLASSINFOpdf
  
\else

\fi

\usepackage{subfigure}
\usepackage{lineno}
\usepackage{listings}

\usepackage{mdwmath}
\usepackage{mdwtab}
\usepackage{multirow}
\usepackage{multicol}
\usepackage{array}
\usepackage{url}
\usepackage{booktabs}

\usepackage{enumitem}
\usepackage{xspace}
\usepackage[export]{adjustbox}
\usepackage{graphicx}
\usepackage{color,soul}
\usepackage{rotating}
\usepackage{setspace}
\usepackage{amsmath} 
\usepackage{amssymb}
\usepackage{float}

\usepackage[colorlinks,
            linkcolor=blue,
            urlcolor=blue,
            anchorcolor=blue,
            citecolor=blue]{hyperref}

\usepackage[textsize=scriptsize,backgroundcolor=yellow!40]{todonotes}

\begin{document}

\title{Design-Pattern-as-a-Service for Blockchain-based Self-Sovereign Identity}

\author{
\IEEEauthorblockN{
Yue Liu\IEEEauthorrefmark{1},
Qinghua Lu\IEEEauthorrefmark{2}\IEEEauthorrefmark{3}\thanks{Qinghua Lu is the corresponding author. Email: qinghua.lu@data61.csiro.au},
Hye-Young Paik\IEEEauthorrefmark{3},
Xiwei Xu\IEEEauthorrefmark{2}\IEEEauthorrefmark{3},
Shiping Chen\IEEEauthorrefmark{2},
Liming Zhu\IEEEauthorrefmark{2}\IEEEauthorrefmark{3}\\
}
\IEEEauthorblockA{\IEEEauthorrefmark{1}College of Computer Science and Technology, China University of Petroleum (East China), China}\\
\IEEEauthorblockA{\IEEEauthorrefmark{2}Data61, CSIRO, Australia}\\
\IEEEauthorblockA{\IEEEauthorrefmark{3}School of Computer Science and Engineering, University of New South Wales, Australia\\
}
}

\markboth{IEEE Software Special Issue on Blockchain and Smart Contract Engineering}%
{Shell \MakeLowercase{\textit{et al.}}: Bare Demo of IEEEtran.cls for Computer Society Journals}

\IEEEtitleabstractindextext{
\begin{abstract}
Self-sovereign identity (SSI) is considered to be a ``killer application" of blockchain. However, there is a lack of systematic architecture designs for blockchain-based SSI systems to support methodical development. An aspect of such gap is demonstrated in current solutions, which are considered coarse grained and may increase data security risks. In this paper, we first identify the lifecycles of three major SSI objects (i.e., key, identifier, and credential) and present fine-grained design patterns critical for application development. These patterns are associated with particular state transitions, providing a systematic view of system interactions and serving as a guidance for effective use of these patterns. Further, we present an SSI platform architecture, which advocates the notion of Design-Pattern-as-a-Service. Each design pattern serves as an API by wrapping the respective pattern code to ease application development and improve scalability and security. We implement a prototype and evaluate it on feasibility and scalability. 

\end{abstract}

}

\maketitle

\IEEEdisplaynontitleabstractindextext

\IEEEpeerreviewmaketitle

\IEEEraisesectionheading{\section{Introduction}\label{sec:introduction}}
\IEEEPARstart{A} legal identity of an entity (i.e. an individual/organization) is defined as a set of attributes about the entity (e.g., name). Identity management includes maintaining identity attribute data and controlling their access, which is fundamentally required in the digitized world. There are three key roles in identity management: \textit{holder}, \textit{issuer}, and \textit{verifier}. 
An entity who registers an identifier associated with some identity attribute data in a particular system is considered as an identity \textit{holder}. A credential is a verifiable claim against some identity attribute data (e.g., date of birth) and facts (e.g., 
transcripts) about the holder, which is attested and digitally signed by a credential \textit{issuer}. A credential \textit{verifier} is an entity who requests a specific credential from a trusted issuer and verifies the credential's authenticity through the issuer's signature. 

\begin{figure*}[t]
\begin{center}
\includegraphics[width=\textwidth]{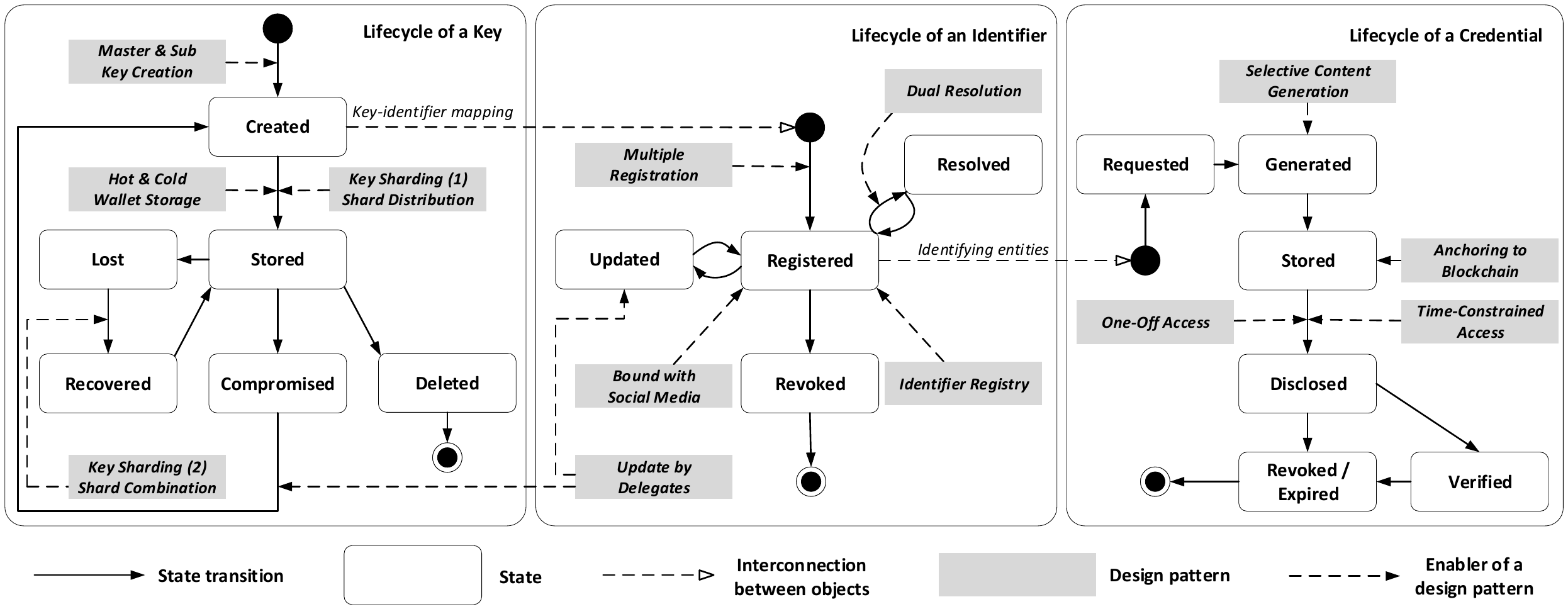}
\caption{Lifecycles of three main objects in Self-Sovereign Identity.} 
\label{lifecycle}
\end{center}
\end{figure*}

Most of the existing identity management solutions require a centralized authority, either for identity registration or credential verification. For example, X.509 certificate~\cite{X509} requires a trusted authority to maintain the mapping between the name and corresponding public key. PGP~\cite{PGP}, as a more decentralized solution, needs to rely on ICANN to assign domain names. Thus, holders often do not have full control over their identity data which may be disclosed or compromised without the holder's knowledge, e.g., Aadhaar data leak~\cite{AadhaarDataLeak}.

Self-sovereign identity (SSI) is an emerging identity management paradigm that allows entities to  have true ownership of their identity data and control their use without involving any intermediary. As self-sovereign identity needs no intermediary, which is aligned with the design nature of blockchain, it is considered to be one of the ``killer applications" of blockchain technology~\cite{Swan:blockchain}. W3C has recently published a design guideline that utilizes blockchains to implement SSI, in which blockchain provides a decentralized public key infrastructure to uniquely register entities via W3C decentralized identifiers (DIDs) without any intermediary~\cite{W3CDID}. Identity credentials can be designed following W3C verifiable credentials data model~\cite{W3CCredential}.  Many organizations are currently exploring how to leverage blockchain to build SSI solutions, e.g., uPort~\cite{uport}, Sovrin~\cite{Sovrin},  Blockstack~\cite{Blockstack}. There are also academic works on the design and development of SSI applications~\cite{grather2018blockchain, indyKYC, takemiya2018sora}. Soltani et al. build a self-sovereign identity framework for customer on-boarding on blockchain \cite{indyKYC}. Takemiya and Vanieiev proposes a blockchain-based protocol for storing encrypted personal information and sharing verifiable claims~\cite{takemiya2018sora}. However, there is a lack of systematic architecture design for blockchain-based SSI systems, which is demonstrated in the currently available solutions. Also, current solutions are considered coarse grained and may lead to less scalable implementation or a reduced level of data security, e.g. inadequate access control for credentials.

Therefore, in this paper, we first identify the lifecycles of three major SSI objects (i.e., key, identifier, and credential), in which the state transitions are associated with fine-grained design patterns that we consider critical to SSI application development. The lifecycles, along with the design pattern annotations, provide an SSI-focused systematic view of system interactions and a guide to effective usage of the design patterns to improve data security and system scalability. Further, based on the design patterns and lifecycles, we present an SSI platform architecture that supports the idea of Design-Pattern-as-a-Service (DPaaS). The platform consists of a full range of SSI services, from key management to credential verification, underwritten by two classes of operations, namely \textit{regular services} and \textit{design pattern services}. Particularly, in each design pattern service, we encapsulate and wrap the respective design pattern code as an API so that the design principles and benefits of the pattern can be easily incorporated into application development. Finally, we implement a proof-of-concept prototype and evaluate it in terms of feasibility and scalability. The results show that our approach is feasible and has scalable performance.

\section{Lifecycles of SSI Objects}
\label{sec:patterns}

Fig. \ref{lifecycle} illustrates three lifecycles of the main objects and their interconnection in SSI: lifecycle of \textit{key}, \textit{identifier} (DID), and \textit{credential}. The figure also highlights our SSI design patterns associating them with relevant states and transitions within the lifecycles. Due to space constraints, we describe the lifecycles below, but introduce pattern details in ~\cite{SSIpattern}. 

\begin{itemize}
\item \textbf{Lifecycle of a Key} The lifecycle of a key starts when a key pair is created for an entity. Once created, it can be stored anywhere the entity prefers. If the private key is lost or compromised, it might be recovered later or replaced with a new private key. The key pair could also be deleted if the entity does not need it anymore.
\item \textbf{Lifecycle of an Identifier} Registering a DID on blockchain using a created key is the beginning of a DID's lifecycle. Each DID is mapped to a blockchain account's public key to ensure uniqueness, and linked to a DDO (DID Document) which specifies the public keys, authentication protocols, and service endpoints. 
Entities establish trust relationships by resolving each involved entity's DID, obtaining its respective DDO and verifying the identity of the entity using the public key kept in DDO. Once a DID is registered, its details contained in the associated DDO can be updated. A DID could also be revoked, if its owner does not need the respective identity anymore.
\item \textbf{Lifecycle of a Credential} The start of a credential's lifecycle is triggered by the request sent by a verifier. The holder receives the request and sends credential requirements to an issuer. The issuer generates the required credential and signs on it. The holder receives the credential and discloses it to the verifier. The verifier can assure the credential authenticity by checking the signature of the issuer. The credential could expire after which it cannot be accessed by the verifier, or be revoked by an issuer if it does not comply with the respective regulations anymore. 

\end{itemize}

\begin{figure*}[t]
\begin{center}
\includegraphics[width=0.85\textwidth]{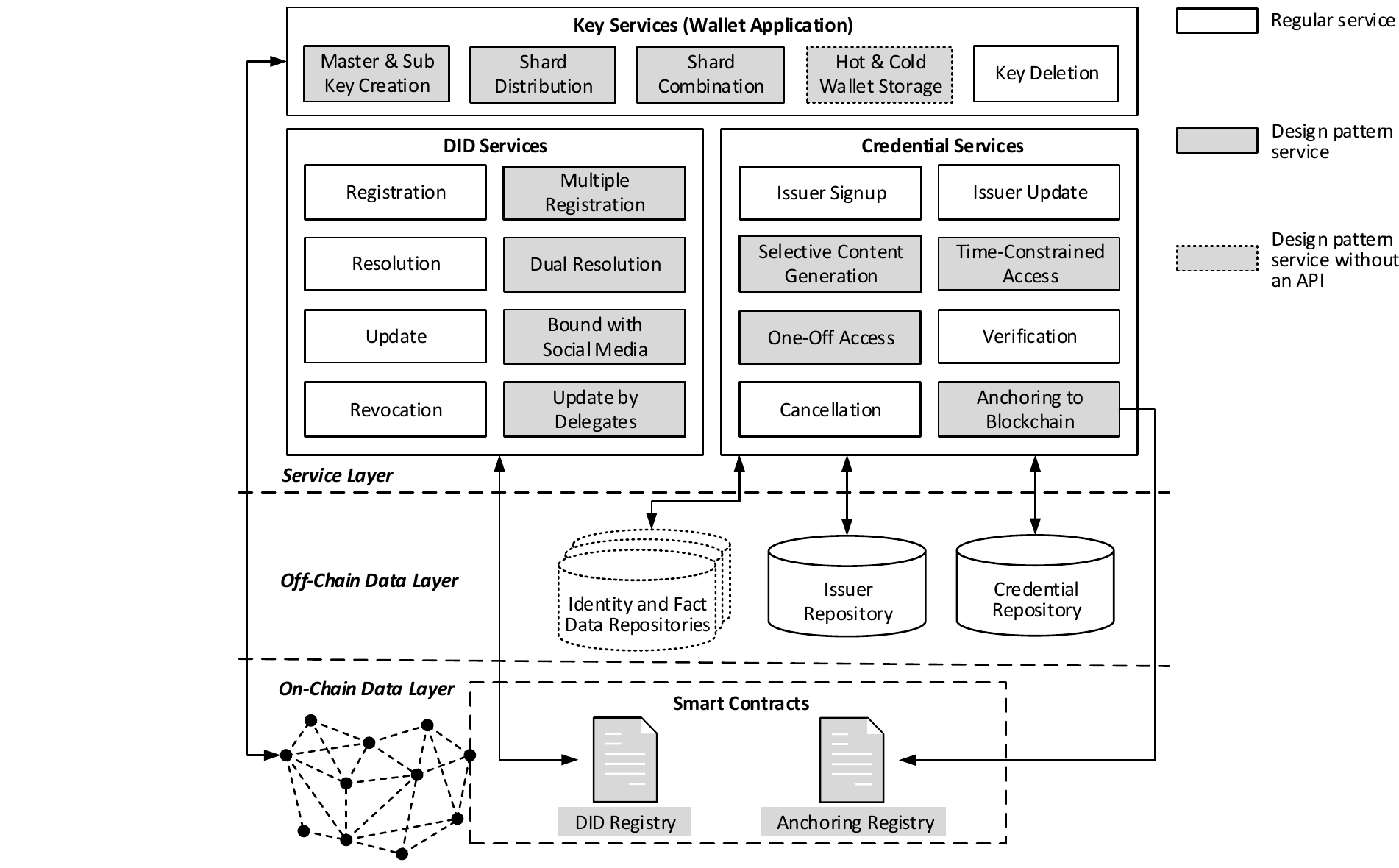}
\caption{Architecture of Design-Pattern-as-a-Service Platform for Self-Sovereign Identity.}
\label{architecture}
\end{center}
\end{figure*}

\section{Design-Pattern-as-a-Service for SSI}
\label{sec:architecture}

In this section, we present a Design-Pattern-as-a-Service (DPaaS) platform architecture towards SSI. Fig.~\ref{architecture} illustrates the overall three-layer architecture: 1) Service Layer, 2) Off-Chain Data Layer, 3) On-Chain Data Layer. The Service Layer is comprised of key services, DID services, and credential services. Services are classified into regular and design pattern services. The code for each design pattern discussed above is implemented, wrapped as an API and delivered as services to facilitate the system development and improve security and scalability.

\subsection{Key Services}
Key management related services are provided in wallet applications. A user (i.e., an entity) is required to create at least one blockchain account which contains a key pair for sending/signing transactions and registering a DID. 

To protect privacy and avoid key compromising, multiple key pairs can be created by  \textit{Master \& Sub Key Generation} to split the management key (i.e., master-key) and signing keys (i.e., sub-keys). A master-key is used for creating/managing all the identities owned by an entity, while sub-keys are used to sign transactions for each identity account representing different identity. Once a sub-key is compromised, one can update the key/DID mapping with a new sub-key using the master-key. 

To recover a lost private key and avoid losing control over the respective blockchain account, \textit{Shard Distribution} breaks the input private key into user-defined number of pieces, while \textit{Shard Combination} rebuilds the private key if there are more input key pieces than the predefined threshold. 

Once a key pair is created, users can choose to store the key in online wallet applications (i.e., hot wallets) or manually write down on a piece of paper (i.e., cold wallet) via \textit{Hot \& Cold Wallet Storage}. Compared to hot wallets, cold wallets are more secure in that the disconnection from the Internet protects them from adversarial online access, but they are less convenient to use. A key pair can be removed via \textit{Key Deletion} when it is not needed anymore.

\subsection{DID Services}
Each DID in the platform is associated with one blockchain account's key pair to ensure its uniqueness. A user can register a DID using one's blockchain account via \textit{Registration} or register multiple DIDs through \textit{Multiple Registration}. \textit{Multiple Registration} allows each key pair to register a DID for every identity an entity has to avoid privacy disclosure caused by transactions correlation. 

To ensure identity data integrity and address blockchain's scalability issue, a lightweight \textit{Identifier Registry} is created as a smart contract on blockchain, which only maintains the mappings between registered DIDs and their associated descriptions (i.e. DDOs). The identity data and facts about a holder are stored in off-chain \emph{Identity and Fact Data Repositories} hosted by the respective issuers (since we focus on legal identities). A registered DID can be \textit{Bound with Social Media} by creating 1-to-1 mapping record in its DDO. This association improves the trustworthiness of both social media profile and blockchain-based identity involved.

\begin{figure*}[t]
	\centering
    \includegraphics[width=0.6\textwidth]{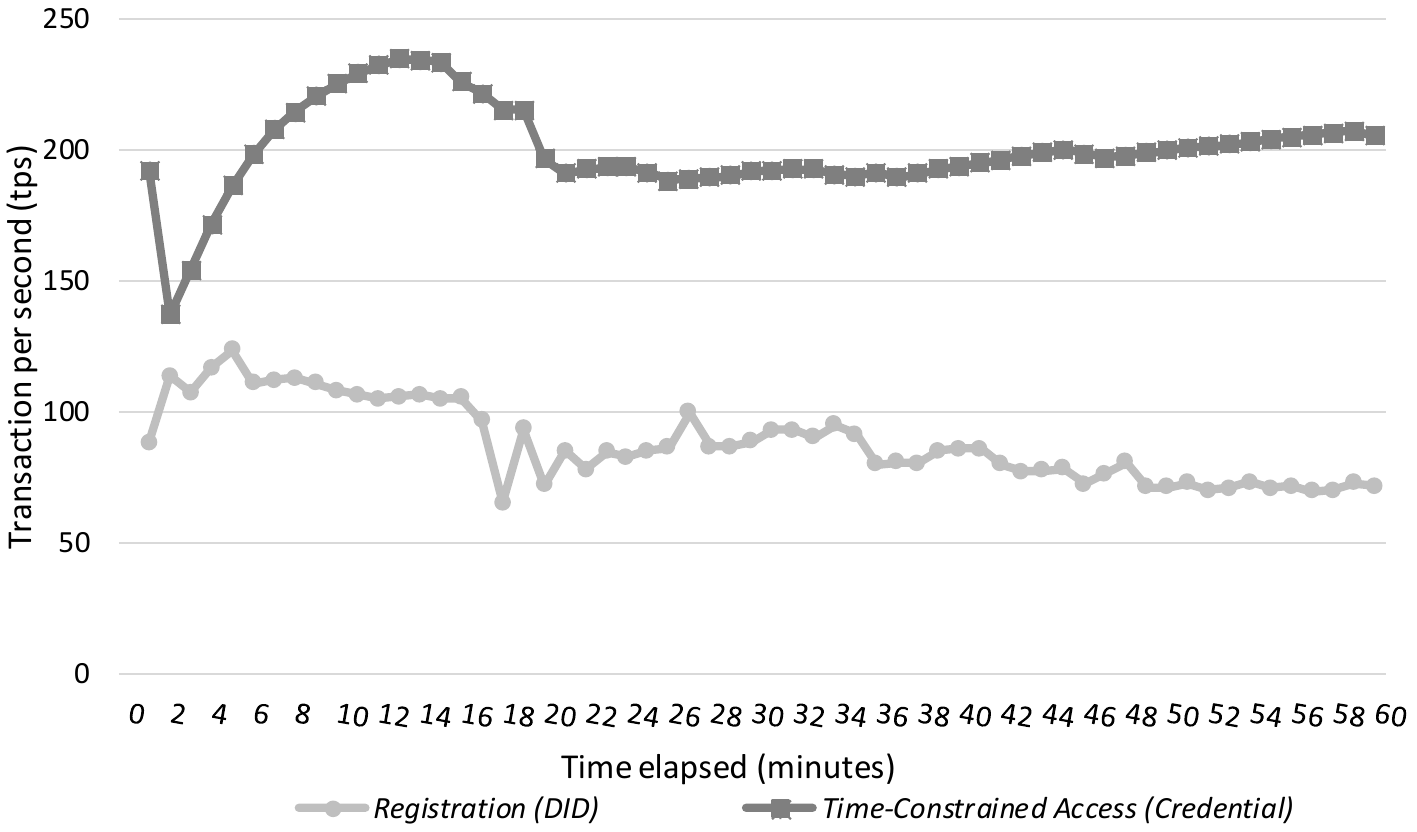}
	\caption{Performance evaluation.}
	\label{throughput}
\end{figure*}

One can obtain a DID's respective public key and service endpoints by acquiring the DDO via \textit{Resolution}. \textit{Dual Resolution} allows two parties to establish a mutual trust relationship with a single operation by extending \textit{Resolution}. A user inputs the DIDs of two involved parties and invokes \textit{Dual Resolution} through which the two parties obtain each other's DDO for resolution. To avoid manipulating DDO, only the DID owner is allowed to update the respective DDO using \textit{Update}. 

When the private key of a user's blockchain account is compromised, the update delegates preregistered in \textit{DID Registry} can vote to replace the DID's compromised key with a new safe key via \textit{Update by Delegates}. A DID can be revoked (e.g., an issuer organization is dissolved) by deleting the associated DDO and marking the DID as revoked in the \textit{DID registry} smart contract via \textit{Revocation}.

\subsection{Credential Services}
Credentials are maintained via credential services. To become a valid issuer, an entity needs to first register via \emph{Issuer Signup}, which requires a super account's (i.e., the platform owner) approval. All the certified issuers' DIDs are maintained in \emph{Issuer Repository} and can be updated by the super account via \emph{Issuer Update}. This design could be further adapted in a decentralized way, by having a group of super issuers register and maintain a new issuer.

To generate a new credential with higher level of security, a holder needs to first send the required credential content requirements to the issuer's service endpoint. The issuer then invokes \textit{Selective Content Generation}, fetching relevant data from the local \textit{Identity and Fact Data Repository} and generating a signed credential in JSON format. The platform generates a JSON Web Token (JWT) and sends it to the issuer once the credential is stored in \textit{Credential Repository}. Each JWT is used to identify and access a credential. Both the credential and JWT are sent by the issuer to the holder's service endpoint.

Once the credential and JWT are received, the holder can share them with the respective verifier. To avoid malicious data disclosure, a holder can limit the accessible period of a credential via \textit{Time-Constrained Access} or \textit{One-Off Access}. A JWT is generated for the credential based on the accessible period. Both of the credential and JWT are added as a new record to \textit{Credential Repository}. When a holder uses \textit{One-Off Access}, the access state flag is updated after the credential is accessed. The holder sends the credential with chosen access period and its respective JWT to the verifier who then verifies it via \textit{Verification}. The platform decodes JWT and checks whether the access period has expired and access flag is false. If the access period is valid, both authenticity and integrity of the credential are checked for verification using the input credential and signature. 

An issuer can revoke an existing credential via \emph{Cancellation} if the holder no longer meets the prerequisites. All revoked/expired credentials cannot be accessed. The integrity of data stored in \textit{Credential Repository}, \textit{Identity and Fact Data Repositories}, and \textit{Issuer Repository} is ensured through \emph{Anchoring to Blockchain}, which periodically stores the hash value of the data in the respective off-chain repository in \emph{Anchoring Registry} smart contract.

\section{Implementation and Evaluation}
\label{sec:evaluation}
We developed a prototype using Node.jsv.10, and wrapped the services into RESTful APIs via express.js. We selected MySQL5.7.17 as the database and Parity1.9.2 with Proof-of-Authority as blockchain. The block gas limit is 80M and the block interval is 5s. Smart contracts are in Solidity with compiler v.0.4.20. 

To evaluate performance, we measure the throughput of \textit{Registration} and \textit{Time-Constrained Access}. We deployed the prototype on Alibaba Cloud (4 vCPUs, 16GB RAM, 40GB disk). The API requests are set to 20 calls/batch and produced via JMeter for 1h. Fig.~\ref{throughput} illustrates that \emph{Registration} is about 70tps, while \emph{Time-Constrained Access} stays around 200tps. The changes in performance over the duration may be caused by the load generator. The fixed block interval and limited block capacity cause an inclusion delay each block which affects the throughput of \emph{Registration}. The performance of \textit{Selective Content Generation} and \emph{One-Off Access} is similar to \emph{Time-Constrained Access}. Also, we tested \emph{Anchoring to Blockchain} with 1000 credentials for 1000 times. The response time is around 0.284s. Overall, the results indicate that our platform achieves scalable performance for organization-level adoption.

\ifCLASSOPTIONcaptionsoff
  \newpage
\fi


\end{document}